%% file: main.tex
\documentclass[twocolumn]{aastex63}
\usepackage{amsmath}
\usepackage{amssymb}
\usepackage{mathrsfs}
\usepackage{makecell}
\usepackage{newtxtext,newtxmath}
\usepackage{xspace}
\usepackage{multirow}

\submitjournal{ApJ}
\shorttitle{Magnetic Fields}
\shortauthors{Farrell et al.}

\definecolor{orange}{rgb}{0.075, 0.533, 0.015}
\newcommand\E[1]{{\color{orange} \bf #1}} 
\newcommand\new[1]{{\color{black}}} 

\input{aliases}
 
\begin{document}

\title{The Initial Magnetic Field Distribution in AB Stars}

\correspondingauthor{Eoin Farrell}

\author[0000-0001-5631-5878]{Eoin Farrell}
\affiliation{Center for Computational Astrophysics, Flatiron Institute, New York, NY 10010, USA}

\author[0000-0001-5048-9973]{Adam S. Jermyn}
\affiliation{Center for Computational Astrophysics, Flatiron Institute, New York, NY 10010, USA}

\author[0000-0002-8171-8596]{Matteo Cantiello}
\affiliation{Center for Computational Astrophysics, Flatiron Institute, New York, NY 10010, USA}
\affiliation{Department of Astrophysical Sciences, Princeton University, Princeton, NJ 08544, USA}

\author[0000-0002-9328-5652]{Daniel Foreman-Mackey}
\affiliation{Center for Computational Astrophysics, Flatiron Institute, New York, NY 10010, USA}

\begin{abstract}
Stars are born with magnetic fields, but the distribution of their initial field strengths remains uncertain.
We combine observations with theoretical models of magnetic field evolution to infer the initial distribution of magnetic fields for AB stars in the mass range $1.6-3.4 M_{\odot}$.
We tested a variety of distributions with different shapes and found that a distribution with a mean of $\sim$ 800 G and a full-width of $\sim$ 600 G is most consistent with the observed fraction of strongly magnetized stars as a function of mass.
Our most-favoured distribution is a Gaussian with a mean $\mu = 770\,\mathrm{G}$ and standard deviation $\sigma = 146\,\mathrm{G}$.
Independent approaches to measure the typical field strength suggest values closer to 2 - 3 kG, a discrepancy which could suggest a mass-dependent and bi-modal initial field distribution, or an alternative theoretical picture for the origin of these magnetic fields.
\end{abstract}


\input{intro}

\input{methods}

\input{IFD}

\input{uncertain}

\input{discuss}

\acknowledgments

The Flatiron Institute is supported by the Simons Foundation.
We are grateful to Gregg Wade for helpful conversations on the IFD.




\appendix

\input{mesa}

\section{Posterior distributions} \label{appendix_posteriors}

\begin{figure*} \centering
\includegraphics[width=\linewidth]{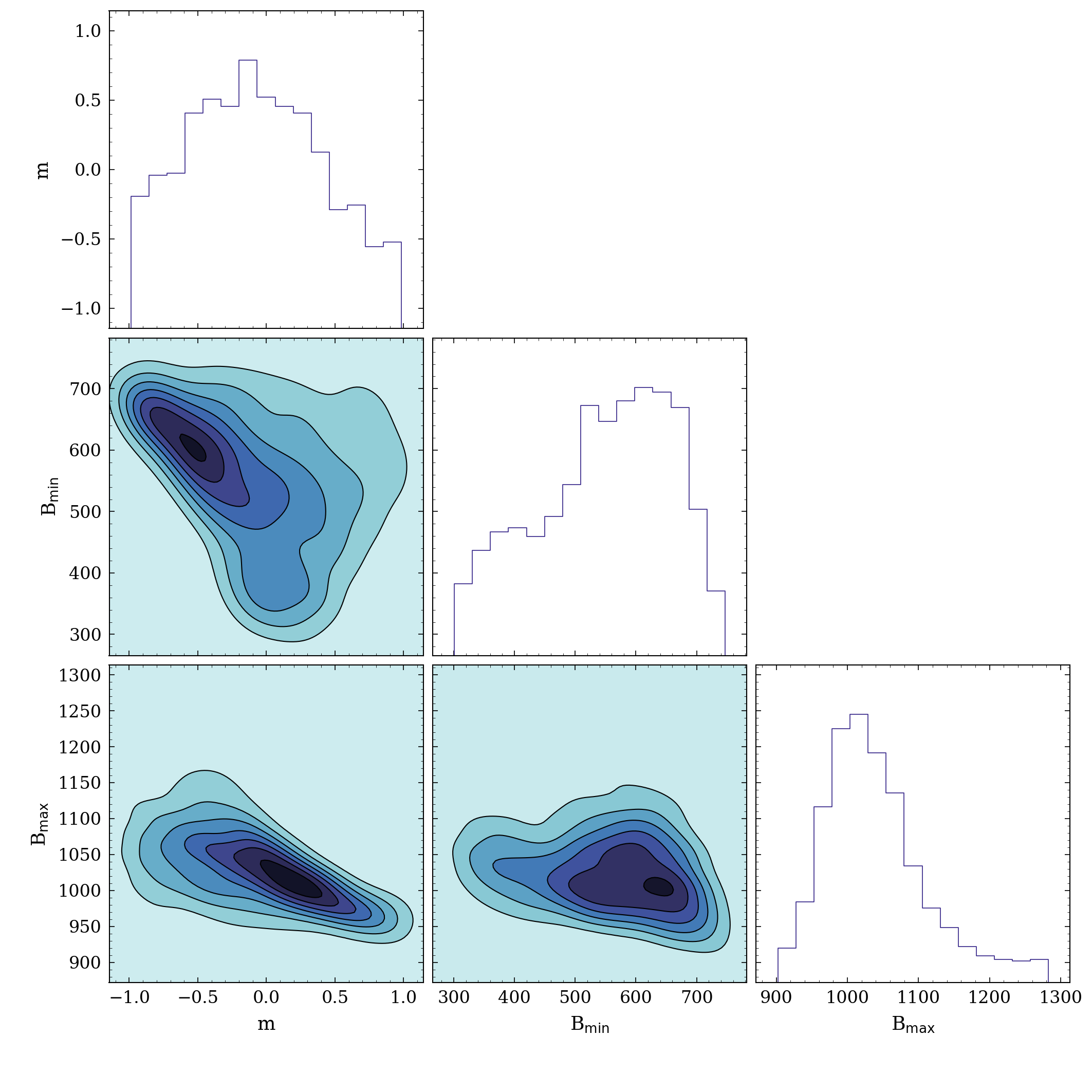}
\caption{Posterior distribution of the slope $m$, the minimum B field value and maximum B field value for the trapezoidal initial field distribution.}
\label{trapezoidal_posterior}
\end{figure*}

\begin{figure*} \centering
\includegraphics[width=\linewidth]{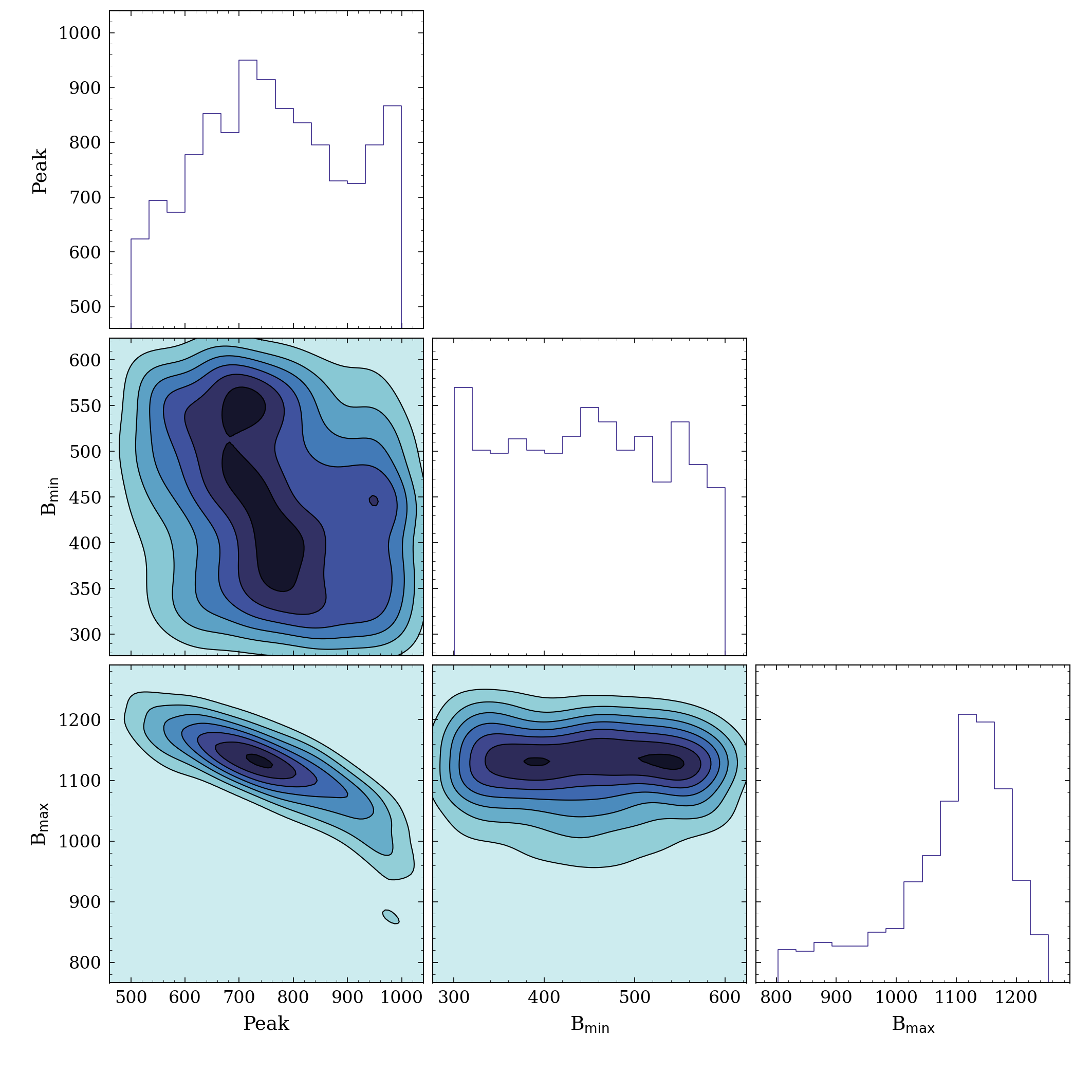}
\caption{Posterior distribution of the location of the peak of the distribution, the minimum B field value and maximum B field value for the triangular initial field distribution.}
\label{triangular_posterior}
\end{figure*}

Figs.~\ref{trapezoidal_posterior} and~\ref{triangular_posterior} show the posterior distributions of the free parameters of the IFD assuming respectively trapezoidal and triangular functional forms.

\bibliography{main_refs, refs}
\bibliographystyle{aasjournal}

\end{document}

%% file: aliases.tex
\newcommand{\msun}{\ensuremath{\mathrm{M}_{\odot}}\xspace}

\newcommand{\teff}{\ensuremath{T_{\rm eff}}\xspace}

\newcommand{\ly}{$L$\xspace}
\newcommand{\bcrit}{\ensuremath{B_{\rm crit}}\xspace}

%% file: intro.tex
\section{Introduction}

Magnetic fields play a significant role in many aspects of stellar evolution.
They transport angular momentum \citep{Spruit2002,Fuller:19}.
They affect stellar winds \citep{Weber1967, Ud-Doula2009}, which in turn can alter planetary evolution \citep{Vidotto2013}.
Magnetic fields can also alter heat transport and produce star spots \citep{Cantiello2011}, influence accretion \citep{Bouvier2007}, and enhance \citep{Harrington2019} or inhibit \citep{Gough1966} chemical mixing.

Stars are expected to form with a range of initial magnetic field strengths.
Unfortunately this Initial Field Distribution (IFD) is difficult to directly observe, as magnetic fields interact with convective regions and can be erased by subsurface convection layers as a star evolves \citep[e.g.][]{Gough1966, Jermyn2020}.
Nonetheless, observations of early-type stars (OBA stars) can probe the present-day distribution of magnetic field strengths, and have revealed that it is \emph{bimodal}.
Stars are observed either with strong fossil fields, in excess of $\sim 300\rm G$, or ultra-weak fields, with amplitudes below a $\sim 3 \rm G$, and these two modes appear to be mass-dependent.
Few or no stars are observed with field strengths in between \citep{Auriere2007, Grunhut2017, Fossati2015}.

The bimodal field distribution suggests that there are at least two different origin stories for early-type stellar magnetism.
Strong magnetic fields may be the relic of fossil field generation during star formation \citep[e.g.][]{Donati:2009} or during a stellar merger \citep{Schneider:16,Schneider:19}, while weak magnetic fields could be the result of ongoing dynamo processes \citep{Cantiello2011,Cantiello2019,Jermyn2021}.

Exploring this scenario,~\cite{Jermyn2020} suggested that the difference between the two modes could be explained by the strength of the initial magnetic field in a star.
Strong enough initial magnetic fields can suppress near-surface convective motions and evolve just under flux conservation, producing the strong-field mode of the present-day distribution.
Below some critical field strength, however, convection begins and can act to erase the near-surface magnetic field, producing the observed weak-field mode.
This naturally explains both modes of the distribution as well as the approximate (mass-dependent) field strengths of the magnetic desert between the modes.

If this suggestion holds then whether or not a star will be observed in the strong-field mode can be calculated given its initial magnetic field strength, mass and evolutionary state.
Hence, the fraction of strongly magnetized stars as a function of mass should depend only on the IFD.
Importantly, this fraction has recently been observed in a volume-limited survey of AB stars by \citet{Sikora2019, Sikora2019a}.

Our aim is to infer the Initial Field Distribution.
We begin in Section~\ref{sec:methods} by introducing the observational data (Section~\ref{sec:obs}), our forward model of the magnetic field evolution (Section~\ref{sec:theory}) and the Approximate Bayesian Computation approach (Section~\ref{sec:abc}) that we use to infer the IFD from the observations.
We then apply these methods and obtain the IFD in Section~\ref{sec:IFD}.
We find that a distribution with a mean of $\sim$ 800 G and a full-width of $\sim$ 600 G is most consistent with the observations, though we cannot precisely infer the shape of the distribution.
We explore uncertainties in our analysis in Section~\ref{sec:uncertainties} and discuss the implications of this IFD in Section~\ref{sec:discuss}.

%% file: methods.tex
\section{Methods}\label{sec:methods}

\input{observations}

\input{theory}

\input{abc}

%% file: observations.tex
\subsection{Observations}\label{sec:obs}

\begin{figure*} \centering
\includegraphics[width=\linewidth]{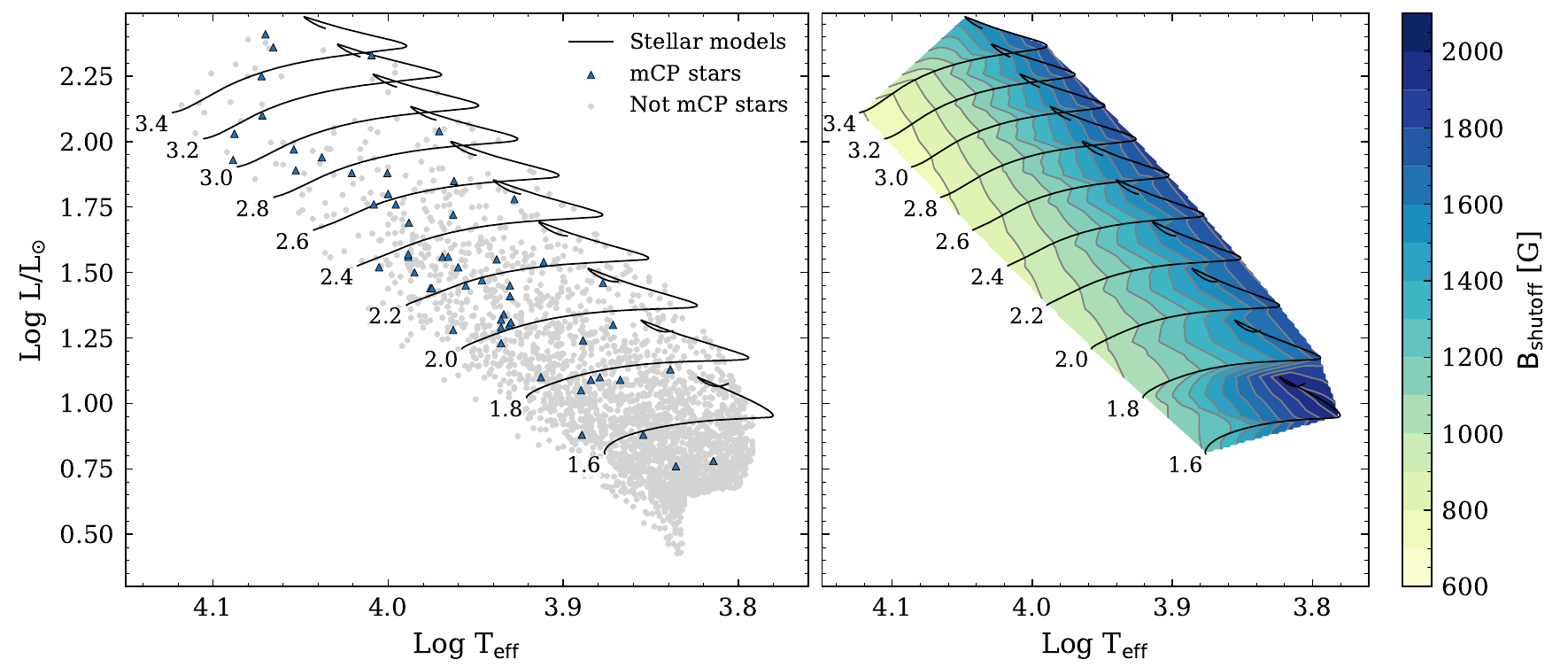}
\caption{\textit{Left:} A Hertzsprung–Russell diagram with observed sample \citep{Sikora2019, Sikora2019a} and our MESA stellar evolution models over-plotted with the mass indicated at the beginning of each track. Non-magnetic AB stars are plotted in grey and strongly-magnetized stars are shown in red. \textit{Right:} The critical magnetic field strength computed using Equation~\ref{eq_bcrit}.}
\label{fig:hrd}
\end{figure*}

\citet{Sikora2019} recently produced a volume-limited survey of all identified intermediate mass MS stars within a heliocentric distance of 100~pc as determined using Hipparcos parallaxes. 
\citet{Sikora2019a} used spectropolarimetry to measure the magnetic field strengths of the magnetic chemically peculiar stars (mCP) in the sample.
Figure~\ref{fig:hrd} shows their sample in the Hertzsprung–Russell (HR) diagram.

\new{We infer the mass of each star in this survey by comparing the value of \ly and \teff reported by \citet{Sikora2019}
to our stellar evolution models, described below in Section~\ref{sec:theory}.}
We then choose to divide the data into five mass bins.
Using less than five bins limits the power of our inference method, while more than five leads to limited sample sizes of just 1 -- 2 stars in the upper mass bins.
In each bin we compute the fraction of strongly-magnetized stars.
We identify objects as strongly magnetic if~\citet{Sikora2019a} were able to measure a magnetic field and as weak otherwise, associating these with the two modes of the present-day field distribution.
Table~\ref{table1} summarises the number of magnetic stars n$_{\rm B}$, non-magnetic stars n$_{\rm nonB}$ and fraction of magnetic stars f$_{\rm B}$ = n$_{\rm B}$/n$_{\rm nonB}$ in the five mass bins.
These five values of f$_{\rm B}$, along with the mass bins, form the data  which we use to infer the IFD.
\new{We emphasise here that our subsequent inference of the initial field distribution is based on the presence or absence of strong magnetic fields as a function of mass, and not on the values of the detected fields.}

\bgroup
\def\arraystretch{1.2}
\begin{table}
\centering
\caption{Characteristics of stars in the volume limited survey by \citet{Sikora2019, Sikora2019a} binned by mass as derived from our stellar evolution models.}
\label{table1}
\begin{tabular}{ r | r | r | r }
\hline
Mass bin (\msun) & \makecell[cr]{Number \\ of mCP stars}  & Total Number & $f_{\rm B}$ \\ 
\hline

1.44 - 1.87 & 10 & 1384 & 0.007 \\ 
1.87 - 2.30 & 19 & 544 & 0.034 \\ 
2.30 - 2.72 & 13 & 192 & 0.068 \\ 
2.72 - 3.15 & 7 & 53 & 0.132 \\ 
3.15 - 3.58 & 6 & 26 & 0.231 \\ 

\hline

\end{tabular}
\end{table}
\egroup

%% file: theory.tex
\subsection{Theory} \label{sec:theory}

\new{Our analysis is based on the overall picture for the evolution of fossil magnetic fields proposed by \cite{Jermyn2020}, which can be summarised as follows.
All stars are born with a large-scale magnetic field.
If this initial magnetic field is stronger than some critical field strength \bcrit, near-surface convective motions can be suppressed and the magnetic field can evolve solely under flux conservation and slow Ohmic diffusion. These produce the observed population of stars with strong magnetic fields.
However, if the magnetic field strength is below the critical field strength (either at the ZAMS or later during the evolution), convection is able to twist magnetic field lines leading to reconnections and the removal of the large-scale surface component of the field.
Convection can still drive weak dynamo-driven surface fields of $\sim 1 G$.
Note that stable fossil fields can still survive underneath the convective regions~\citep{Jermyn2021}, but they are not detectable at the surface.
\cite{Jermyn2020} propose that the observed bimodal field distribution can naturally be explained by this scenario.
}

\new{To implement this picture in our simulations, we}
begin by noting that there is a critical vertical magnetic field strength above which systems are \emph{stable} to convection.
\citet{MacDonald2019} compute this as
\begin{equation} \label{eq_bcrit}
B_{\rm crit}^2 =  \frac{4 \pi \rho c^{2}_{s} Q (\nabla_{\rm rad} - \nabla_{\rm ad})}{1 - Q (\nabla_{\rm rad} - \nabla_{\rm ad}) + d \ln \Gamma_1 / d \ln p},
\end{equation}
where $c_s$ is the sound speed, $\Gamma_1$ is the first adiabatic index, $\nabla_{\rm ad}$ is the adiabatic temperature gradient, $\nabla_{\rm rad}$ is the radiative temperature gradient, $\rho$ is the density, and 
\begin{equation}
Q = \frac{4 - 3 p_{\rm gas}/p}{p_{\rm gas}/p}.
\end{equation}
Here $p_{\rm gas}$ is the gas pressure and $p$ is the total pressure.

Using this criterion, we \new{implement the scenario described by~\citet{Jermyn2020} to}
model the evolution of the magnetic fields assuming the following:
\begin{enumerate}
\item The magnetic field is uniform near the surface of a star.
\item So long as $B > B_{\rm crit}$, convection is suppressed and the field evolves according to flux conservation ($B \propto R^{-2}$, where $R$ is the stellar radius).
\item If the magnetic field is ever $B < B_{\rm crit}$, near-surface convective motions begin and quickly erase the surface magnetic field.
\end{enumerate}

We calculated stellar evolution tracks for stars ranging from $1.6-3.4\,M_\odot$ using revision 15140 of the Modules for Experiments in Stellar Astrophysics
\citep[MESA,][]{Paxton2011, Paxton2013, Paxton2015, Paxton2018, Paxton2019} software instrument.
Details on the MESA microphysics inputs are provided in Appendix~\ref{appen:mesa}.
We swept the mass range in increments of $0.2 M_\odot$ and used solar metallicity ($Z=0.02$).
The evolutionary tracks of our models are plotted in the HR diagram in Fig.~\ref{fig:hrd}. 
In these calculations we forced the temperature gradient to be $\nabla_{\rm rad}$ in the outer envelope where subsurface convective layers would otherwise form.
This allows us to simulate the inhibition of convection by a strong magnetic field in those regions. 
This choice has a very small effect on the surface properties, as shown by \citet{Jermyn2020}.
However, it does impact the thermodynamic properties of the subsurface convection regions, which are relevant for computing \bcrit.

We evaluate $B_{\rm crit}$ for each model at each point in time using equation~\eqref{eq_bcrit} and taking the maximum value inside the subsurface convective layers, because in our scenario even a small convective region is enough to erase the surface magnetic field.
For the stars observed by \citep{Sikora2019} the relevant subsurface convective regions are driven by hydrogen (H~CZ) and helium ionization (HeI~CZ, HeII~CZ) \citep{Cantiello2019}.
We note that the HeI CZ is frequently stabilized by viscosity and thermal diffusion~\citep{Jermyn2022}, so formally we should exclude its contribution to \bcrit. However, our models indicate that the HeI CZ never produces the largest \bcrit among available convection zones.

\begin{figure} \centering
\includegraphics[width=\linewidth]{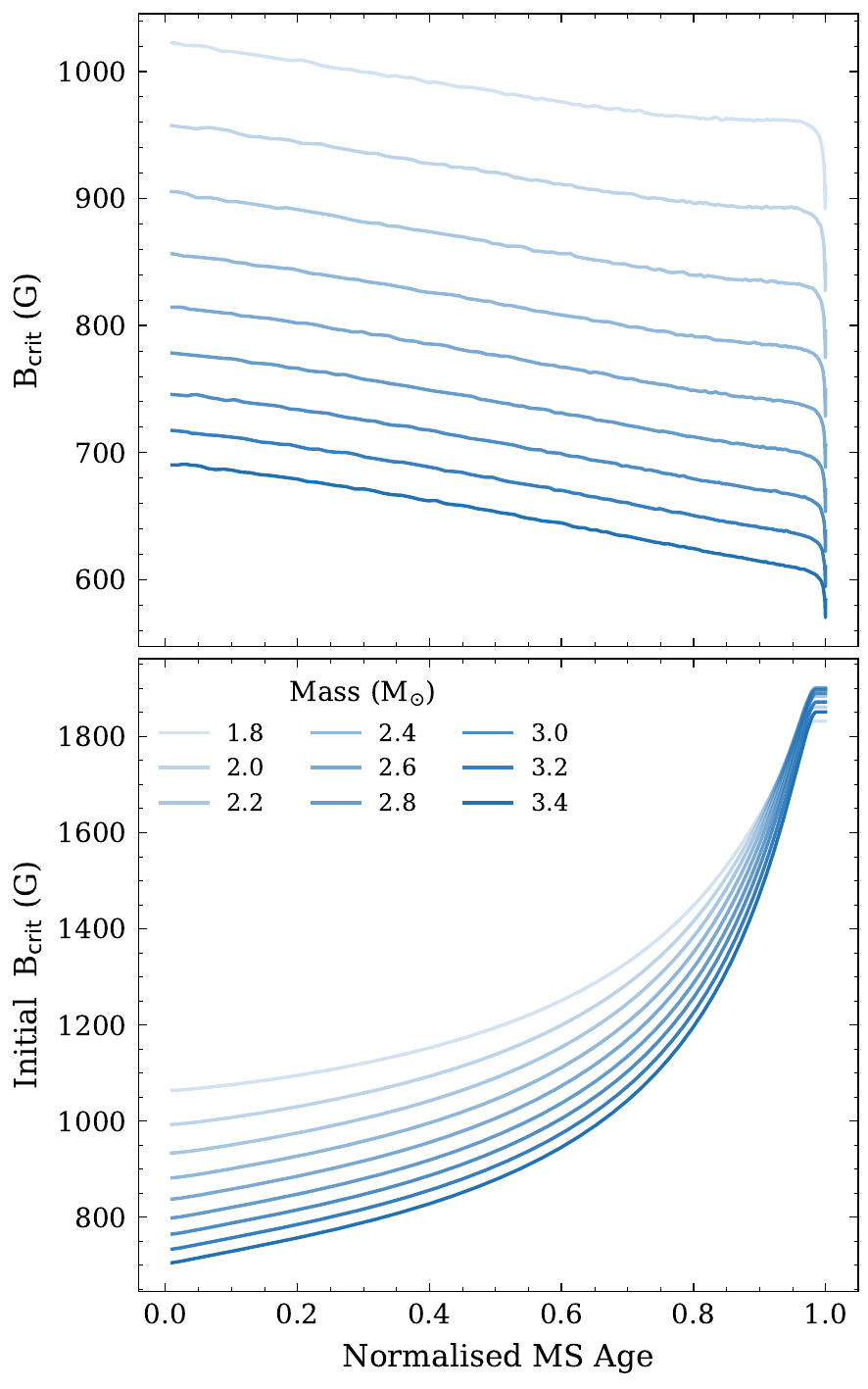}
\caption{\textit{Upper panel}: Critical magnetic field required to prevent the formation of subsurface convective layers as a function of normalised main sequence age for models of different masses. \textit{Lower panel}: The critical initial magnetic field required to suppress the formation of subsurface convective layers and maintain a strong magnetic field until a given age as a function of normalised main sequence age. The trend in age is due to expansion during the main sequence.}
\label{fig:bcrit}
\end{figure}

Fig.~\ref{fig:bcrit} (upper) shows \bcrit as a function of mass and evolutionary stage in our models.
Similar to \citet{Jermyn2020}, we find that the critical field strength decreases with increasing stellar mass.
We also find that forcing the temperature gradient to the radiative gradient in the subsurface convective layers typically decreases \bcrit by about 1 \% relative to what~\citet{Jermyn2020} calculated, though this can increase to 10\% for low-mass models at towards the late main-sequence.

Stars expand during the main-sequence (MS).
Assuming conservation of magnetic flux, the surface magnetic field $B$ decrease with radius $R$ as $B \propto 1/R^2$.
Therefore the initial field strength required to produce a strong (super-critical) magnetic field for a given mass and MS fractional age is the cumulative maximum of \bcrit from the upper panel in Fig.~\ref{fig:bcrit} scaled by the square of the increase of the radius. This \emph{initial} \bcrit is plotted in lower panel of Fig.~\ref{fig:bcrit}, where the increase as a function of MS age is due to the expansion of the star. As in the upper panel, the critical field strength decreases with increasing mass.

%% file: abc.tex
\subsection{Approximate Bayesian Computation}\label{sec:abc}

We use an Approximate Bayesian Computation (ABC) method \citep[e.g.][]{Alsing2018, sunnaker2013} to perform likelihood-free inference of the initial field distribution.
The ABC method allows us to reconstruct the IFD as a probability density function $I(B) = dN/dB$ from the observed fraction of strongly magnetised stars f$_{\rm B}(M_\star)$ using only forward simulations, free from any likelihood assumptions or approximations.
The advantage of using ABC is that it is relatively easy to to simulate mock data based on a given IFD, but it is not feasible to analytically propagate an IFD to an expected distribution of observed fields.
We use the \texttt{elfi} python package \citep{elfi2018} for ABC to perform our simulations, which take place as follows.

First, we choose a parameterised, functional form for $I(B)$.
We write this as $I_{\mu,\sigma,...}(B)$, where the subscripts denote the parameterization.
For simplicity, we assume that the fossil field distribution is mass independent throughout. We discuss the implications of this assumption later on.

Next, we choose an appropriate prior distribution $p(\mu,\sigma,...)$ over these free parameters.
For example, one of our parameterizations is a Gaussian distribution described by a mean $\mu$ and a standard deviation $\sigma$, and we choose flat priors over $\mu$ and $\sigma$.

Finally, we enter a loop:
\begin{itemize}
	\item We sample values for the free parameters of our distribution from the prior.
	\item We construct a population of stars with initial magnetic field strengths sampling the resulting IFD ($I_{...}(B)$). These stars have masses and ages chosen to match those in the observed volume-limited sample described above.
	\item For each star, we use our stellar evolution models and frozen flux evolution to determine if the magnetic field is weaker than the critical field at any point between the pre-main-sequence and the age of the star inferred from observations. If this occurs, we mark that star as `not magnetic', assuming that convection will erase the fossil field. Otherwise we mark the star as `magnetic'.
	\item We bin the simulated stars by mass and compute the magnetic fraction $f_{\rm B}$ in each bin.
	\item We compare $f_{\rm B}$ to that reported from observations by computing the sum of the absolute differences between the fractions in each mass bin.
	\item If this difference is less than some threshold distance, the values of the parameters chosen at the beginning are saved, otherwise they are rejected.
\end{itemize}
We perform this loop 10$^{5}$ times for each functional form of $I(B)$.
The accepted samples form our posterior distribution over the free parameters of each functional form.
We verified that the accepted posterior distributions did not change substantially when we increased the number of iterations in the loop or the threshold distance between the simulations and observations

%% file: IFD.tex
\section{Inferred Initial Field Distribution}\label{sec:IFD}

\bgroup
\def\arraystretch{1}
\begin{table*}
\centering
\caption{Details of a range of distributions that we tested, including the free parameters, prior distributions, means and standard deviations ($\sigma$) of the posterior distributions and the `distance' between the threshold 1000th best simulation and the observations (D$_{1000}$).}
\label{table2}
\begin{tabular}{ l | l | l | r | r | r}

\hline
\rule{0pt}{1ex}
Distribution & Parameters & Priors & Posterior  Mean & Posterior Standard Deviation & D$_{1000}$\\
[1ex]

\hline
\multicolumn{6}{c}{Standard Assumptions} \\
[1ex]
\hline

Linear      &  m (slope)    & U(-1, 1)  &    -0.12      &  0.01 & 0.39 \\

\rule{0pt}{4ex}   \multirow{3}{*}{Trapezoidal} & m (slope)         & U(-1, 1)     & -0.086    & 0.48  &  \multirow{3}{*}{0.07}\\
                             & B$_{\rm min}$    & U(300, 850)  & 551       & 106   &  \\
                             & B$_{\rm max}$    & U(850, 1350) & 1028      & 53    &  \\

\rule{0pt}{4ex}   \multirow{2}{*}{Gaussian}     & $\mu$         & U(500, 1000))     & 770    & 44  &  \multirow{2}{*}{0.06}\\
                                                & $\sigma$      & U(10, 230)        & 146    & 37   &  \\

 \rule{0pt}{4ex}   \multirow{3}{*}{Triangular} & Midpoint         & U(500, 1000)     & 767    & 131  &  \multirow{3}{*}{0.05}\\
                             & B$_{\rm min}$    & U(300, 600)  & 445       & 85   &  \\
                             & B$_{\rm max}$    & U(800, 1300) & 1107      & 79    &  \\

\hline
\multicolumn{6}{c}{Additional Tests} \\
[1ex]
\hline
 
\rule{0pt}{4ex}   \multirow{3}{*}{Gaussian + $f_{0}$} & $\mu$         & U(500, 1000)     & 843    & 64  &  \multirow{3}{*}{0.07}\\
                                                      & $\sigma$    & U(10, 230)  & 115       & 49   &  \\
                                                      & f$_{0}$    & U(0, 1) & 0.34      & 0.20    &  \\

\rule{0pt}{4ex}
Gaussian +               & $\mu_0$         & U(500, 1000)     & 731    & 172  &  \multirow{3}{*}{0.03}\\
mass dependence                           & $\sigma$    & U(10, 230)  & 150       & 47   &  \\
($\mu = \mu_0 + \beta * m$)            & $\beta$    & U(-100, 100) & 14      & 52    &  \\

\hline

\end{tabular}
\end{table*}
\egroup

Because magnetohydrodynamics is complicated, we do not have a strong physical prior for the functional form of the IFD. 
For simplicity, we begin by assuming that all stars are born with some non-zero magnetic field and that the distribution is independent of stellar mass.
We discuss the impact of relaxing these assumptions later on. 
We tested the following different forms for the IFD, all of which are summarized in the upper half of Table~\ref{table2}.
\begin{enumerate}
\item Linear: A distribution over $B\in [300,1500]\mathrm{G}$ of the form $I(B) = I_0 + m B$, where $I_0$ is determined to normalize the distribution. We assign the slope $m$ a uniform prior distribution over $[-1,1]$.
\item Trapezoidal: A variant of the linear distribution but with variable left and right endpoints, resulting in two additional free parameters ($B_{\rm min}$, $B_{\rm max}$), which we assign broad uniform priors.
\item Gaussian: A Gaussian distribution characterized by a mean $\mu$ and variance $\sigma^2$. Both $\mu$ and $\sigma$ are given uniform priors.
\item  Triangular: A distribution with a single peak, dropping to zero linearly on either side. Both the peak location and the left and right roots are allowed to vary, with all three having uniform priors.
\end{enumerate}

\begin{figure*} \centering
\includegraphics[width=0.98\linewidth]{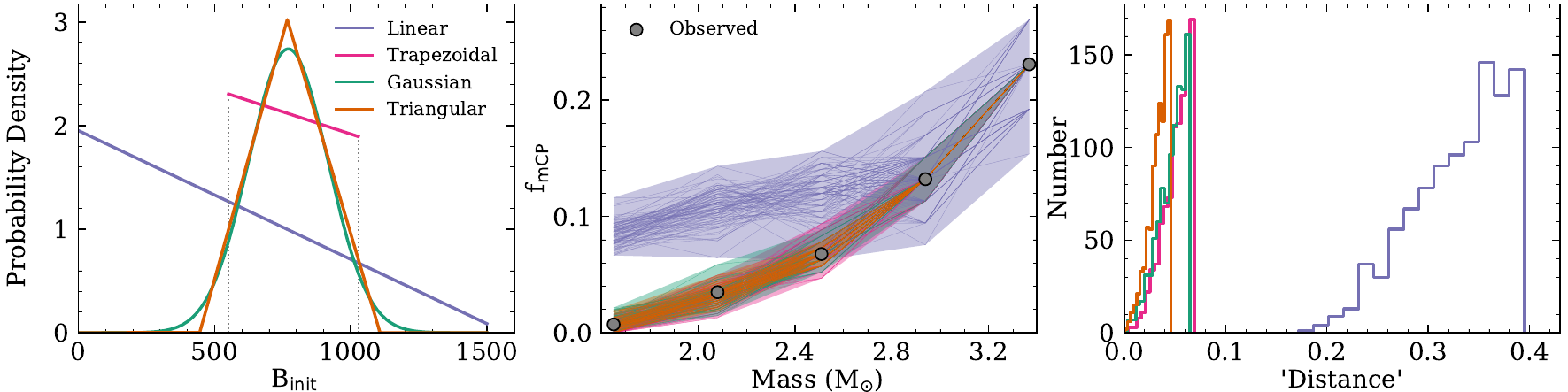}
\caption{Comparing the results from the ABC simulations for different functional forms of the fossil field distribution. \textit{Left panel:} Plots of the fossil field distributions with the free parameters equal to the means of their posterior distributions. \textit{Middle panel:} Comparison between outputs of simulations and observations. \textit{Right panel:} The distribution of the ``distance'' between the observations and simulations for each distribution type.}
\label{compare_functions}
\end{figure*}

To properly compare these different forms we performed our ABC procedure over $10^5$ distributions of each form, sampling the distribution parameters from their priors.
We chose our acceptance tolerance separately for each form so as to accept the best $10^3$ samples, so that the posterior would be well-sampled, and used these to construct the posterior distributions for each parameter.
We verified that the posterior distributions we find are not sensitive to this acceptance threshold.
Corner plots of these posteriors are provided in Appendix~\ref{appendix_posteriors}.

The left panel of Fig.~\ref{compare_functions} shows examples of all four forms, with their respective parameters set to the means of their respective posterior distributions.
The middle panel then compares the distributions from the 100 best simulations for each functional form with the observations.
Finally, the right panel shows the distribution of the ``distance'' between the observations and simulations for each distribution type. 
The ‘distance’ is the sum of the distance between each of the observational data points in the middle panel and the corresponding simulated points.

We find that a linear distribution from 300 G to 1500 G is inconsistent with the observed data (Fig.~\ref{compare_functions}, middle). 
The distribution of distances in the lower panel of Fig.~\ref{compare_functions} confirms that linear distributions perform poorly compared with the other distributions. 
The reason for this is that the linear distribution produces many more strongly-magnetized $\sim 1.8 \msun$ stars than are actually observed.
By contrast, the trapezoidal, Gaussian and triangular forms are all able to reproduce the observations within a much tighter tolerance, and visually reproduce the observed trend in $f_{\rm B}$ much more closely (Fig.~\ref{compare_functions}, middle).

The trapezoidal form is a useful one for building intuition about our inference procedure.
This form favours a relatively uniform distribution from 500 G to 1100 G.
Because the critical magnetic field strength decreases with increasing stellar mass, the lower bound of the distribution is sensitive to higher mass stars and the upper bound of the distribution is sensitive to lower mass stars.
The upper bound of the trapezoidal distribution is relatively strongly constrained due to the very low fraction of strongly magnetised $\sim\,1.8 \msun$ stars.
The lower bound and slope are less constrained because the fraction of strongly magnetised stars at higher masses just tells us how many stars have field strengths below the lowest $B_{\rm crit}$, not how those stars are distributed.

The Gaussian and triangular forms are very similar in shape, and both rather different from the trapezoidal form. 
They both favour a peak around $800 \rm G$ and a similar width as that of the trapezoidal distribution (e.g. $\sigma \sim 130\rm G$).
This suggests that the initial field distribution could have a peak at 800 G and a full-width of around 600 G, with very few stars born with magnetic fields of less than 500 G or greater than 1100 G.

Given that Gaussian, triangular, and trapezoidal all fit the data similarly well we cannot tell much else about the shape of the distribution.
At the same time, the fact that all three forms obtain similar IFDs is encouraging, and tells us that features which all three agree on, like the mean and the width of the IFD, are likely robust.

The trapezoidal and triangular forms, with their sharp kinks and discontinuities, seem less likely to be found in nature than the Gaussian
Therefore, given the similarity between their performance and that of the Gaussian, we choose to explore the Gaussian form in more detail below.

\begin{figure} \centering
\includegraphics[width=\linewidth]{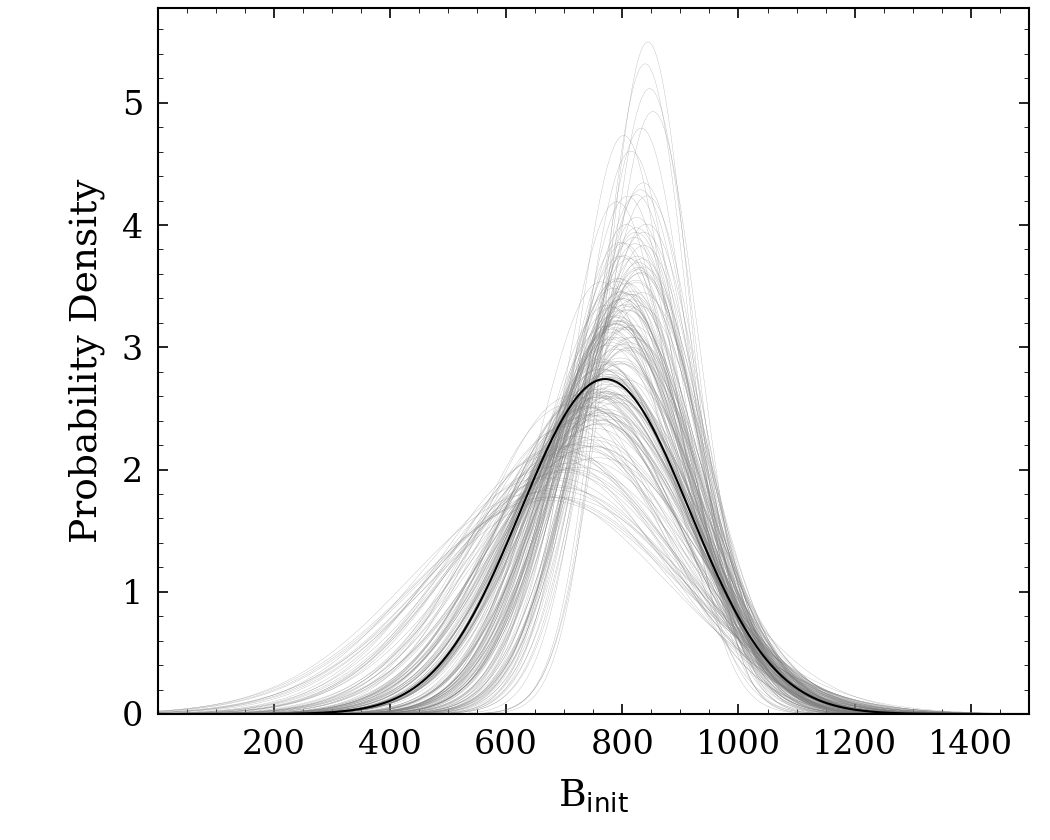}
\caption{The black line indicates the IFD using the means from the posterior distributions. A selection of samples from the posterior distribution over Gaussian IFDs are shown in grey.}
\label{gaussian_cloud}
\end{figure}

The posterior distribution for the Gaussian form has mean $\mu = 770 \pm 44\,\mathrm{G}$ and standard deviation $\sigma = 145 \pm 77\,\mathrm{G}$.
These are anti-correlated, such that we see that distributions with higher mean tend to be narrower, and those with higher variance tend to have lower means.
Fig.~\ref{gaussian_cloud} shows samples of distributions drawn from this posterior, as well as the average over these samples, which broadly tells the same story as above: the initial field distribution is peaked around 800 G and has a full-width of around 600 G, with very few stars born below 500 G or above 1100 G.

\begin{figure} \centering
\includegraphics[width=\linewidth]{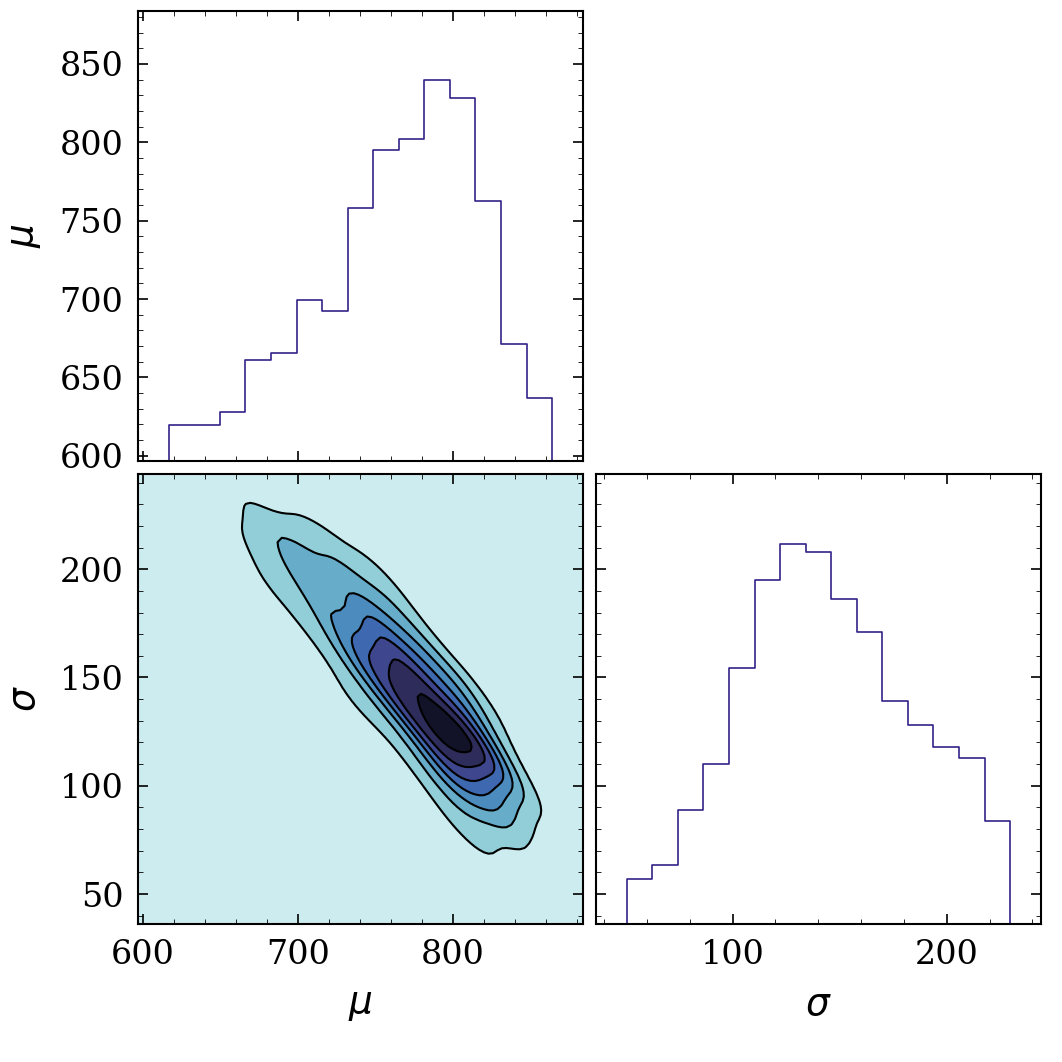}
\caption{Posterior distributions for the mean $\mu$ and the standard deviation $\sigma$ are shown for the Gaussian form.}
\label{gaussian_posteriors}
\end{figure}

%% file: uncertain.tex
\section{Uncertainties}\label{sec:uncertainties}

We now consider various uncertainties in our approach.

\subsection{Bimodal IFDs} \label{bimodal}

We have assumed that all stars were born with some non-zero initial magnetic field. 
While this is almost certainly the case, it could well be that the IFD is itself bimodal, and observations of young stars suggest the possibility of a second mode at very weak field strengths~\citep{2019A&A...622A..72V}.

\begin{figure*} \centering
\includegraphics[width=\linewidth]{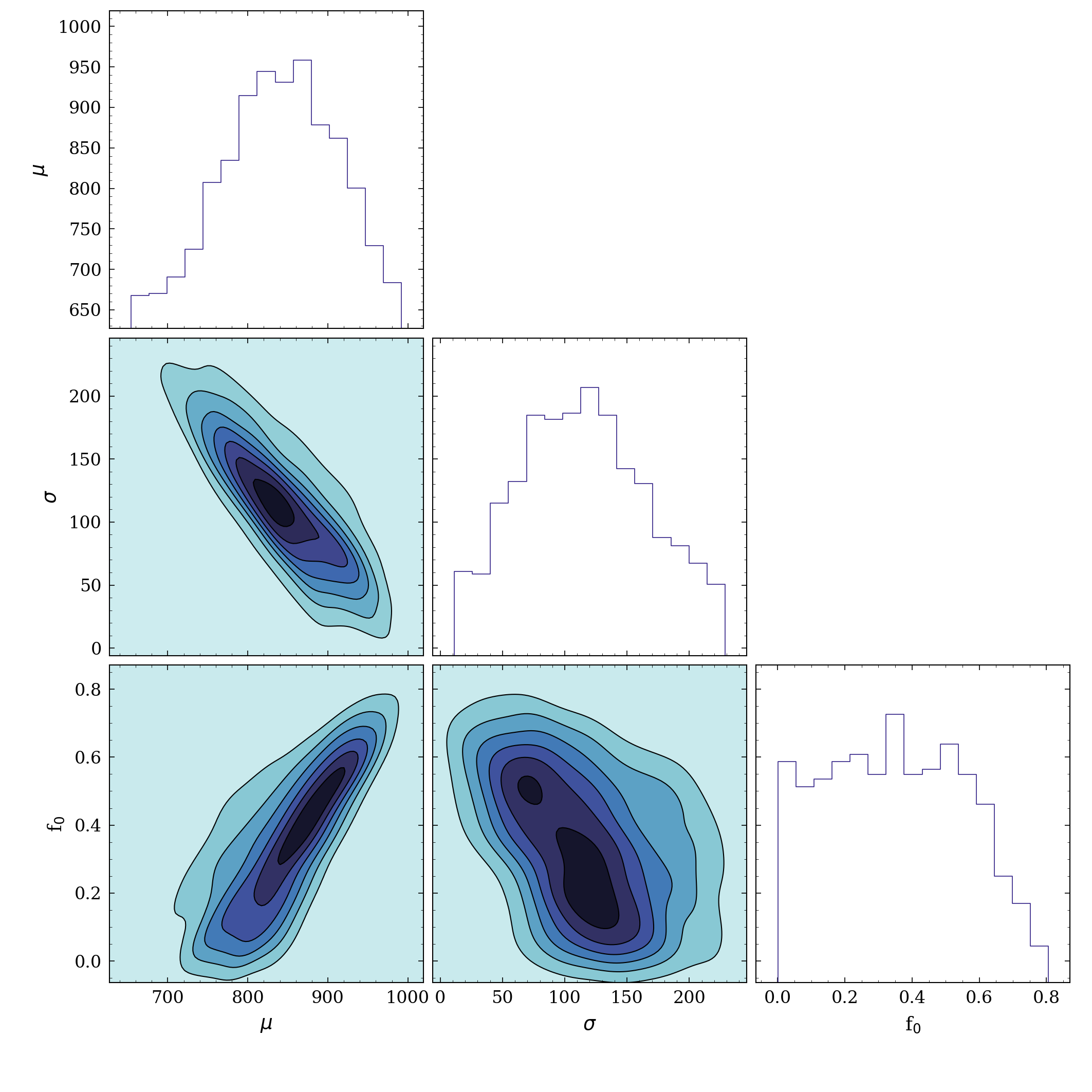}
\caption{Posterior distribution of the mean $\mu$ and standard deviation $\sigma$ and fraction of stars born with no strong magnetic field ($f_0$) for a Gaussian initial field distribution.}
\label{gaussian_w0_posterior}
\end{figure*}

To explore this possibility we investigated distributions in which some fraction of stars are born with an initial magnetic field of $B_{\rm init} = 0$, while the rest have initial fields distributed according to a Gaussian form.
We refer to this as the Gaussian+$f_0$ form.
Fig.~\ref{gaussian_w0_posterior} shows the posterior distribution over $f_0$, $\mu$, and $\sigma$ for this form.
We see that while $\sigma$ is mostly uncorrelated with $f_0$, $\mu$ and $f_0$ are strongly correlated.
This makes sense: as the fraction of stars with zero initial field strength increases, more stars with strong magnetic fields are needed to match the observed magnetic fraction $f_{\rm B}$.
For this reason, regardless of the precise form we choose, any functional form allowing an additional mode at weak field strengths will exhibit such a degeneracy.
We cannot exclude such forms, and this degeneracy introduces an additional uncertainty to our models.


\subsection{Possible Mass Dependence}
\begin{figure*} \centering
\includegraphics[width=\linewidth]{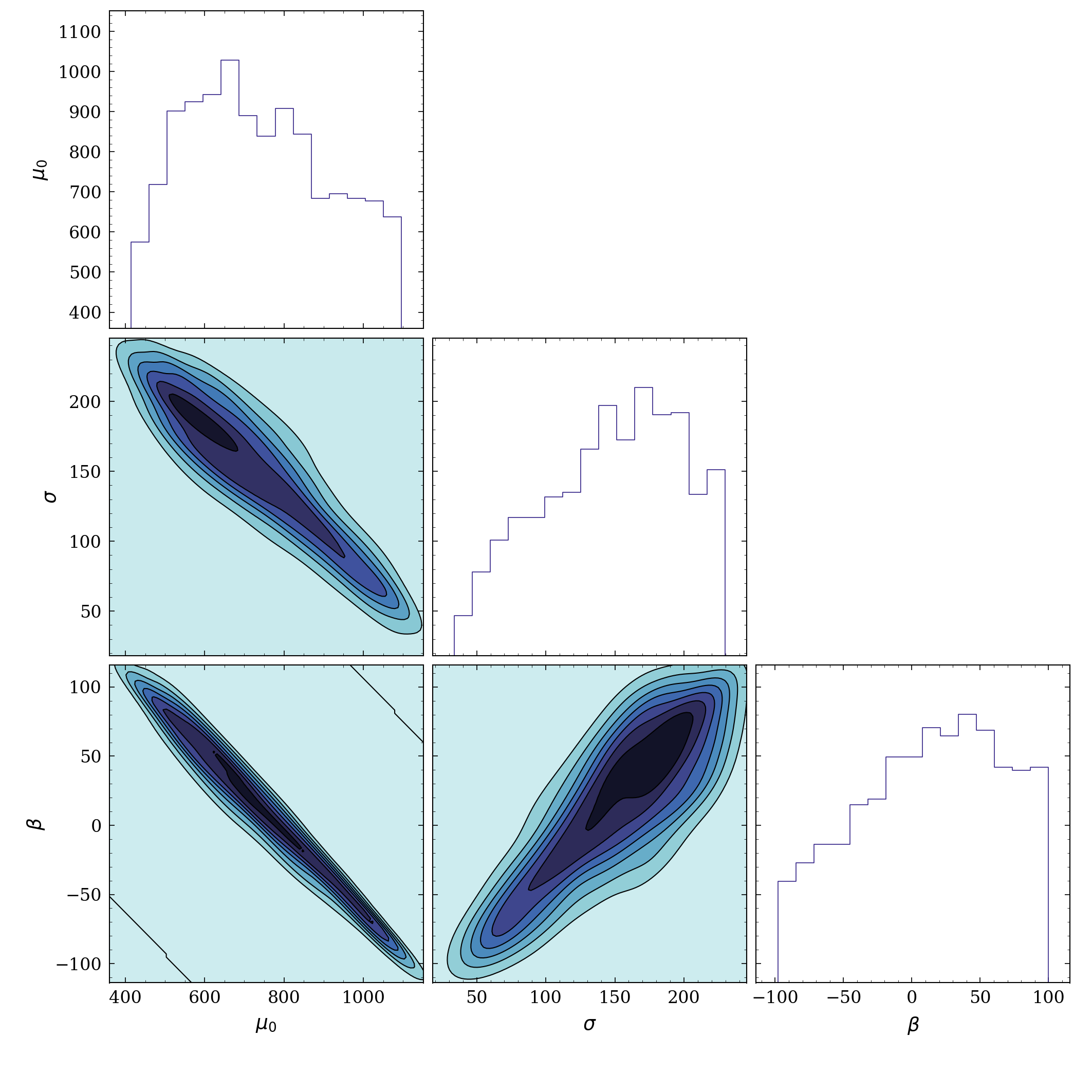}
\caption{Posterior distribution of IFD parameters for a Gaussian form and allowing a dependence of the mean of the Gaussian with the mass through $\mu = \mu_0 + \beta * m$.}
\label{gaussian_mass_posterior}
\end{figure*}

We investigated the possibility of a dependence of the distribution of initial fossil field on mass. 
To do this, we tested a Gaussian IFD with a mean that scales linearly with the stellar mass $m$ such that $\mu = \mu_0 + \beta * m$.
We found that $\mu_0$ and $\beta$ are strongly degenerate (Fig.~\ref{gaussian_mass_posterior}).

To understand this, note that our approach is primarily using the fact that \bcrit varies with mass, and so differences in the magnetic fraction between different mass bins tell us about the IFD.
As a result if we permit our parameterized IFD to vary with mass we effectively lose our ability to constrain the distribution.
What little inferential power we retain comes from the evolution of \bcrit across the main sequence, but we do not have a large enough sample of stars to also bin by age and so this is only minimally constraining.
As a result, we emphasize that our results are strongly contingent on the assumption of a mass-independent IFD over the range from $1.6-3.4\msun$.

\new{
\citet{Sikora2019a} derived the current distribution of magnetic field strengths and found a log-normal distribution centered at 2.6$\,$kG and extending to about 10$\,$kG. 
However, the Gaussian IFD that we derive predicts a current distribution centered around 1$\,$kG, with only a small spread.
One possible explanation for this discrepancy is that the IFD is mass-dependent and bi-modal, with a component of stars that are born with $B_{\rm init} = 0$.
Such an IFD could allow stronger fields to develop, while also allowing most of the weaker fields to be erased by convection.
The discrepancy between the field strengths observed by \citet{Sikora2019a} and the IFD that we derive could suggest such mass-dependence and bi-modality.
However, as we discussed above, our inference method does not allow us to constrain a mass-dependent and bi-modal IFD.
Another possible explanation is that there is a problem with our assumptions about the interaction of magnetic fields and convection, which would negate our picture of the initial field distribution.
}





\subsection{Sensitivity to overshooting}

To study the sensitivity of our results to the details of our stellar evolution calculations, we computed a grid of models with a higher value of core convective overshooting ($f_{\rm ov} = 0.020$ compared to $f_{\rm ov} = 0.014$). Convective overshooting affects the size of stellar cores, and consequently the stellar luminosity and lifetime.  We find that the posterior distributions of the parameters inferred using stellar models with larger overshooting is consistent with the distribution inferred using the original models to within 1\%.

\subsection{Magnetic Field Evolution}

We have assumed that, so long as the magnetic field is strong enough to shut off convection,  the field strength evolves according to flux conservation from the beginning of the main sequence phase onward.  This neglects a variety of different possible phenomena. On the main sequence the most important are magnetic diffusion, differential rotation, and stellar interactions.

We think it is safe to neglect magnetic diffusion, as the diffusion time across a star is long compared with its main-sequence lifetime \citep[e.g.][]{Braithwaite:2017}.
We likewise suspect that differential rotation can be neglected, as super-critical magnetic fields are likely strong enough to inhibit any significant shears from developing on the main-sequence.

Stellar interactions, on the other hand, do pose a challenge to our analysis. Mass accretion can potentially enhance surface stellar magnetic fields \citep{Grunhut:2013}, and very strongly magnetized stars are believed to be the outcome of stellar mergers \citep{Schneider:2019}.  Depending on the incidence of stellar interactions on the main sequence, these processes could impact our ability to infer the IFD at the ZAMS. While the expected fraction of massive stars undergoing interactions during the main sequence is fairly high \citep[$\sim$ 30\%,][]{Mink:2014}, it is much smaller for the mass range explored here. This is because the fraction of short period binaries is smaller in these later-type stars \citep{Moe:2017}. In the future it would be useful to carefully quantify the impact of stellar interactions on the magnetization of main-sequence, intermediate mass stars.

%% file: discuss.tex
\section{Discussion}\label{sec:discuss}

We have inferred the initial distribution of magnetic field strengths (IFD) from a volume-limited sample of AB stars with magnetic field measurements.
To do this we have assumed that the IFD is independent of stellar mass and that convection erases near-surface subcritical magnetic fields.
Our favoured distribution is a Gaussian with mean $\mu = 770\,\mathrm{G}$ and standard deviation $\sigma = 145\,\mathrm{G}$.
We now turn to the astrophysical implications of this distribution.

\subsection{Magnetic Desert}

\citet{Jermyn2020} proposed that the observed distribution of magnetic fields - and the ``magnetic desert'' -- can be explained in a scenario where some fraction of stars are born with a smooth distribution of initial fields and these fields are then modified by convection.
The fact that our preferred IFD reproduces the observed magnetic fraction in AB stars is consistent with this hypothesis.

It is important to note that this did not need to be the case: we could not have successfully fit arbitrary data.
For instance this story would have trouble reproducing a non-monotone magnetic fraction as a function of mass, and would need to invoke more complicated assumptions such as a mass-dependent IFD.

\subsection{Connection to Star Formation}

It is unclear what physical mechanism sets the IFD.
It is possible that the IFD reflects some frozen-in flux present in the molecular cloud a star formed from.
However, that story runs into trouble explaining the mean of the IFD.
Typical molecular clouds are have strong enough magnetic fields that they cannot collapse without shedding magnetic energy~\citep{2008ApJ...680..457T}.
If just enough magnetic field is shed to allow the cloud to collapse then we should expect initial magnetic fields in approximate equipartition with gas pressure, predicting field strengths which are orders of magnitude stronger than what is observed.

On the other hand if convection indeed erases any magnetic fields that came before then we would expect the magnetic field on the ZAMS to be generated by a convective dynamo as the star descends the Hayashi track.
\citet{2019A&A...622A..72V} compiled measurements of magnetic fields on the pre-main-sequence in their figure~1.
For fully convective stars at the base of the Hayashi track they see a broad distribution of magnetic field strengths ranging from $\sim 100\,\rm G$ to $\sim 800\,\rm G$.
These field strengths are in equipartition with the kinetic energy of convection in the outer layers of a star carrying luminosity $L \sim L_\odot$, and occur for stars with luminosities of that order, so it is at least plausible that the these fields are dynamo-generated.

From that point on though the story becomes unclear.
These relatively strong fields persist through the development of a radiative interior, though they mostly vanish by the time early-type stars develop convective cores and radiative envelopes~\citep{2019A&A...622A..72V}, so there could well be non-trivial field evolution happening in the immediate evolution prior to the ZAMS.

\subsection{Solar Magnetism}

If the IFD we have inferred holds at lower masses, we should expect the Sun to have been born with a $\sim 800\,\rm G$ magnetic field, and that field could well have survived below the solar convection zone to this day~\citep{1953sun..book..532C}.

So far as we know, the fossil field in the interior of the Sun has not been measured.
There are, however, various upper bounds which have been obtained.
Reasoning from measurements of the solar oblateness,~\citet{2004ApJ...601..570F} found an upper bound of $7\times 10^6\,\rm G$, which is certainly consistent with our prediction.
By contrast ~\citet{1996ApJ...458..832B} inferred an upper bound of $\sim 30\,\mathrm{G}$ in the radiative zone based on the lack of bias in the solar cycle, contingent on some assumptions including that the dynamo is localized to the tachocline.
If their assumptions hold then this tight limit is evidence against our IFD, and favours more complicated models like the bimodal Gaussian+$f_0$ forms we considered in Section~\ref{sec:uncertainties}.

The results of \citet{Stello:2016} also support the idea that the interior of stars like the Sun is not strongly magnetized. 
They used asteroseismology to infer the presence of strong magnetic fields in evolving red giant cores \citep{Fuller:19}, and claimed that the incidence of core magnetization goes to zero for stars below 1.1$M_\odot$. These stars are the descendants of main sequence stars with radiative cores, while our inference used stars with convective cores on the main sequence. The interaction of convective cores with a large scale magnetic field could be an important piece of the puzzle \citep[e.g.][]{Featherstone:2009}.

\subsection{Compact Remnants}

Our results predict that some early-type stars with subcritical surface magnetic fields may have strong fossil fields hiding just beneath their subsurface convection zones~\citep{Jermyn2021}.
Assuming simple flux conservation \new{and that the interior fields are at least as strong as at the surface}, a \new{fossil} field of $\sim$ 800 G would give rise to fields of order $10^6\,\rm G$ in red giant cores and white dwarfs (WD) and $10^{15}\,\rm G$ in neutron stars (NS) \citep[e.g.][]{Woltjer1964, Duncan:1992}.
Asteroseismology has revealed magnetic fields of order $10^6\,\rm G$ in about 20\% of red giant cores~\citep{2015Sci...350..423F,Stello:2016}, and roughly 20\% of white dwarfs have $10^{5-6}\,\rm G$ fields~\citep{2019A&A...628A...1L}. 
This does not necessarily mean that all strongly magnetized NS (magnetars) and magnetic WD inherit their fields directly from the IFD. 
\new{
The fact that the fraction of mCP stars on the MS ($\sim 10\%$) appears to be less than the fraction of magnetic white dwarfs ($\sim 20\%$, \citealt{Bagnulo2021}) and that the distribution of magnetic field strengths in white dwarfs spreads across 4 dex of field strength \citep{Bagnulo2021} suggest that, at the very least, not all magnetic white dwarfs are formed from mCP MS stars.
}
Indeed there are other mechanisms for generating or destroying magnetic fields in the course of stellar evolution~\citep{Duncan:1992,2002A&A...381..923S,2019MNRAS.485.3661F}, but the scenario discussed here provides one possible, simple mechanism for the generation of (strongly) magnetized compact remnants.

%% file: mesa.tex
\section{MESA} \label{appen:mesa}

The MESA EOS is a blend of the OPAL \citep{Rogers2002}, SCVH
\citep{Saumon1995}, FreeEOS \citep{Irwin2004}, HELM \citep{Timmes2000},
and PC \citep{Potekhin2010} EOSes.

Radiative opacities are primarily from OPAL \citep{Iglesias1993,
Iglesias1996}, with low-temperature data from \citet{Ferguson2005}
and the high-temperature, Compton-scattering dominated regime by
\citet{Buchler1976}.  Electron conduction opacities are from
\citet{Cassisi2007}.

Nuclear reaction rates are from JINA REACLIB \citep{Cyburt2010} plus
additional tabulated weak reaction rates \citet{Fuller1985, Oda1994,
Langanke2000}.
Screening is included via the prescription of \citet{Chugunov2007}.
Thermal neutrino loss rates are from \citet{Itoh1996}.

The MESA input files are available as a ZENODO repository: \citet{zenodo_dataset}.

%% file: main.bbl
\begin{thebibliography}{}
\expandafter\ifx\csname natexlab\endcsname\relax\def\natexlab#1{#1}\fi
\providecommand{\url}[1]{\href{#1}{#1}}
\providecommand{\dodoi}[1]{doi:~\href{http://doi.org/#1}{\nolinkurl{#1}}}
\providecommand{\doeprint}[1]{\href{http://ascl.net/#1}{\nolinkurl{http://ascl.net/#1}}}
\providecommand{\doarXiv}[1]{\href{https://arxiv.org/abs/#1}{\nolinkurl{https://arxiv.org/abs/#1}}}

\bibitem[{{Alsing} {et~al.}(2018){Alsing}, {Wandelt}, \& {Feeney}}]{Alsing2018}
{Alsing}, J., {Wandelt}, B., \& {Feeney}, S. 2018, \mnras, 477, 2874,
  \dodoi{10.1093/mnras/sty819}

\bibitem[{{Auri{\`e}re} {et~al.}(2007){Auri{\`e}re}, {Wade}, {Silvester},
  {Ligni{\`e}res}, {Bagnulo}, {Bale}, {Dintrans}, {Donati}, {Folsom},
  {Gruberbauer}, {Hui Bon Hoa}, {Jeffers}, {Johnson}, {Landstreet},
  {L{\`e}bre}, {Lueftinger}, {Marsden}, {Mouillet}, {Naseri}, {Paletou},
  {Petit}, {Power}, {Rincon}, {Strasser}, \& {Toqu{\'e}}}]{Auriere2007}
{Auri{\`e}re}, M., {Wade}, G.~A., {Silvester}, J., {et~al.} 2007, \aap, 475,
  1053, \dodoi{10.1051/0004-6361:20078189}

\bibitem[{{Boruta}(1996)}]{1996ApJ...458..832B}
{Boruta}, N. 1996, \apj, 458, 832, \dodoi{10.1086/176861}

\bibitem[{{Bouvier} {et~al.}(2007){Bouvier}, {Alencar}, {Boutelier},
  {Dougados}, {Balog}, {Grankin}, {Hodgkin}, {Ibrahimov}, {Kun}, {Magakian}, \&
  {Pinte}}]{Bouvier2007}
{Bouvier}, J., {Alencar}, S.~H.~P., {Boutelier}, T., {et~al.} 2007, \aap, 463,
  1017, \dodoi{10.1051/0004-6361:20066021}

\bibitem[{{Braithwaite} \& {Spruit}(2017)}]{Braithwaite:2017}
{Braithwaite}, J., \& {Spruit}, H.~C. 2017, Royal Society Open Science, 4,
  160271, \dodoi{10.1098/rsos.160271}

\bibitem[{{Buchler} \& {Yueh}(1976)}]{Buchler1976}
{Buchler}, J.~R., \& {Yueh}, W.~R. 1976, \apj, 210, 440, \dodoi{10.1086/154847}

\bibitem[{{Cantiello} \& {Braithwaite}(2011)}]{Cantiello2011}
{Cantiello}, M., \& {Braithwaite}, J. 2011, \aap, 534, A140,
  \dodoi{10.1051/0004-6361/201117512}

\bibitem[{{Cantiello} \& {Braithwaite}(2019)}]{Cantiello2019}
---. 2019, \apj, 883, 106, \dodoi{10.3847/1538-4357/ab3924}

\bibitem[{{Cassisi} {et~al.}(2007){Cassisi}, {Potekhin}, {Pietrinferni},
  {Catelan}, \& {Salaris}}]{Cassisi2007}
{Cassisi}, S., {Potekhin}, A.~Y., {Pietrinferni}, A., {Catelan}, M., \&
  {Salaris}, M. 2007, \apj, 661, 1094, \dodoi{10.1086/516819}

\bibitem[{{Chugunov} {et~al.}(2007){Chugunov}, {Dewitt}, \&
  {Yakovlev}}]{Chugunov2007}
{Chugunov}, A.~I., {Dewitt}, H.~E., \& {Yakovlev}, D.~G. 2007, \prd, 76,
  025028, \dodoi{10.1103/PhysRevD.76.025028}

\bibitem[{{Cowling}(1953)}]{1953sun..book..532C}
{Cowling}, T.~G. 1953, in The Sun, ed. G.~P. {Kuiper}, 532

\bibitem[{{Cyburt} {et~al.}(2010){Cyburt}, {Amthor}, {Ferguson}, {Meisel},
  {Smith}, {Warren}, {Heger}, {Hoffman}, {Rauscher}, {Sakharuk}, {Schatz},
  {Thielemann}, \& {Wiescher}}]{Cyburt2010}
{Cyburt}, R.~H., {Amthor}, A.~M., {Ferguson}, R., {et~al.} 2010, \apjs, 189,
  240, \dodoi{10.1088/0067-0049/189/1/240}

\bibitem[{{de Mink} {et~al.}(2014){de Mink}, {Sana}, {Langer}, {Izzard}, \&
  {Schneider}}]{Mink:2014}
{de Mink}, S.~E., {Sana}, H., {Langer}, N., {Izzard}, R.~G., \& {Schneider},
  F.~R.~N. 2014, \apj, 782, 7, \dodoi{10.1088/0004-637X/782/1/7}

\bibitem[{{Donati} \& {Landstreet}(2009)}]{Donati:2009}
{Donati}, J.~F., \& {Landstreet}, J.~D. 2009, \araa, 47, 333,
  \dodoi{10.1146/annurev-astro-082708-101833}

\bibitem[{{Duncan} \& {Thompson}(1992)}]{Duncan:1992}
{Duncan}, R.~C., \& {Thompson}, C. 1992, \apjl, 392, L9, \dodoi{10.1086/186413}

\bibitem[{Farrell {et~al.}(2022)Farrell, Jermyn, Cantiello, \&
  Foreman-Mackey}]{zenodo_dataset}
Farrell, E., Jermyn, A., Cantiello, M., \& Foreman-Mackey, D. 2022, {Supporting
  data for "The Initial Magnetic Field Distribution in AB Stars"},  Zenodo,
  \dodoi{10.5281/zenodo.6139141}

\bibitem[{{Featherstone} {et~al.}(2009){Featherstone}, {Browning}, {Brun}, \&
  {Toomre}}]{Featherstone:2009}
{Featherstone}, N.~A., {Browning}, M.~K., {Brun}, A.~S., \& {Toomre}, J. 2009,
  \apj, 705, 1000, \dodoi{10.1088/0004-637X/705/1/1000}

\bibitem[{{Ferguson} {et~al.}(2005){Ferguson}, {Alexander}, {Allard}, {Barman},
  {Bodnarik}, {Hauschildt}, {Heffner-Wong}, \& {Tamanai}}]{Ferguson2005}
{Ferguson}, J.~W., {Alexander}, D.~R., {Allard}, F., {et~al.} 2005, \apj, 623,
  585, \dodoi{10.1086/428642}

\bibitem[{{Fossati} {et~al.}(2015){Fossati}, {Castro}, {Sch{\"o}ller},
  {Hubrig}, {Langer}, {Morel}, {Briquet}, {Herrero}, {Przybilla}, {Sana},
  {Schneider}, {de Koter}, \& {BOB Collaboration}}]{Fossati2015}
{Fossati}, L., {Castro}, N., {Sch{\"o}ller}, M., {et~al.} 2015, \aap, 582, A45,
  \dodoi{10.1051/0004-6361/201526725}

\bibitem[{{Friedland} \& {Gruzinov}(2004)}]{2004ApJ...601..570F}
{Friedland}, A., \& {Gruzinov}, A. 2004, \apj, 601, 570, \dodoi{10.1086/380480}

\bibitem[{{Fuller} {et~al.}(1985){Fuller}, {Fowler}, \& {Newman}}]{Fuller1985}
{Fuller}, G.~M., {Fowler}, W.~A., \& {Newman}, M.~J. 1985, \apj, 293, 1,
  \dodoi{10.1086/163208}

\bibitem[{{Fuller} {et~al.}(2015){Fuller}, {Cantiello}, {Stello}, {Garcia}, \&
  {Bildsten}}]{2015Sci...350..423F}
{Fuller}, J., {Cantiello}, M., {Stello}, D., {Garcia}, R.~A., \& {Bildsten}, L.
  2015, Science, 350, 423, \dodoi{10.1126/science.aac6933}

\bibitem[{{Fuller} {et~al.}(2019{\natexlab{a}}){Fuller}, {Piro}, \&
  {Jermyn}}]{Fuller:19}
{Fuller}, J., {Piro}, A.~L., \& {Jermyn}, A.~S. 2019{\natexlab{a}}, \mnras,
  485, 3661, \dodoi{10.1093/mnras/stz514}

\bibitem[{{Fuller} {et~al.}(2019{\natexlab{b}}){Fuller}, {Piro}, \&
  {Jermyn}}]{2019MNRAS.485.3661F}
---. 2019{\natexlab{b}}, \mnras, 485, 3661, \dodoi{10.1093/mnras/stz514}

\bibitem[{{Gough} \& {Tayler}(1966)}]{Gough1966}
{Gough}, D.~O., \& {Tayler}, R.~J. 1966, \mnras, 133, 85,
  \dodoi{10.1093/mnras/133.1.85}

\bibitem[{{Grunhut} {et~al.}(2013){Grunhut}, {Wade}, {Leutenegger}, {Petit},
  {Rauw}, {Neiner}, {Martins}, {Cohen}, {Gagn{\'e}}, {Ignace}, {Mathis}, {de
  Mink}, {Moffat}, {Owocki}, {Shultz}, {Sundqvist}, \& {MiMeS
  Collaboration}}]{Grunhut:2013}
{Grunhut}, J.~H., {Wade}, G.~A., {Leutenegger}, M., {et~al.} 2013, \mnras, 428,
  1686, \dodoi{10.1093/mnras/sts153}

\bibitem[{{Grunhut} {et~al.}(2017){Grunhut}, {Wade}, {Neiner}, {Oksala},
  {Petit}, {Alecian}, {Bohlender}, {Bouret}, {Henrichs}, {Hussain},
  {Kochukhov}, \& {MiMeS Collaboration}}]{Grunhut2017}
{Grunhut}, J.~H., {Wade}, G.~A., {Neiner}, C., {et~al.} 2017, \mnras, 465,
  2432, \dodoi{10.1093/mnras/stw2743}

\bibitem[{{Harrington} \& {Garaud}(2019)}]{Harrington2019}
{Harrington}, P.~Z., \& {Garaud}, P. 2019, \apjl, 870, L5,
  \dodoi{10.3847/2041-8213/aaf812}

\bibitem[{{Iglesias} \& {Rogers}(1993)}]{Iglesias1993}
{Iglesias}, C.~A., \& {Rogers}, F.~J. 1993, \apj, 412, 752,
  \dodoi{10.1086/172958}

\bibitem[{{Iglesias} \& {Rogers}(1996)}]{Iglesias1996}
---. 1996, \apj, 464, 943, \dodoi{10.1086/177381}

\bibitem[{{Irwin}(2004)}]{Irwin2004}
{Irwin}, A.~W. 2004, The FreeEOS Code for Calculating the Equation of State for
  Stellar Interiors.
\newblock \url{http://freeeos.sourceforge.net/}

\bibitem[{{Itoh} {et~al.}(1996){Itoh}, {Hayashi}, {Nishikawa}, \&
  {Kohyama}}]{Itoh1996}
{Itoh}, N., {Hayashi}, H., {Nishikawa}, A., \& {Kohyama}, Y. 1996, \apjs, 102,
  411, \dodoi{10.1086/192264}

\bibitem[{{Jermyn} {et~al.}(2022){Jermyn}, {Anders}, \&
  {Cantiello}}]{Jermyn2022}
{Jermyn}, A.~S., {Anders}, E.~H., \& {Cantiello}, M. 2022, arXiv e-prints,
  arXiv:2201.10567.
\newblock \doarXiv{2201.10567}

\bibitem[{{Jermyn} \& {Cantiello}(2020)}]{Jermyn2020}
{Jermyn}, A.~S., \& {Cantiello}, M. 2020, \apj, 900, 113,
  \dodoi{10.3847/1538-4357/ab9e70}

\bibitem[{{Jermyn} \& {Cantiello}(2021)}]{Jermyn2021}
---. 2021, \apj, 923, 104, \dodoi{10.3847/1538-4357/ac2d2a}

\bibitem[{{Landstreet} \& {Bagnulo}(2019)}]{2019A&A...628A...1L}
{Landstreet}, J.~D., \& {Bagnulo}, S. 2019, \aap, 628, A1,
  \dodoi{10.1051/0004-6361/201936009}

\bibitem[{{Langanke} \& {Mart{\'{\i}}nez-Pinedo}(2000)}]{Langanke2000}
{Langanke}, K., \& {Mart{\'{\i}}nez-Pinedo}, G. 2000, Nuclear Physics A, 673,
  481, \dodoi{10.1016/S0375-9474(00)00131-7}

\bibitem[{Lintusaari {et~al.}(2018)Lintusaari, Vuollekoski,
  Kangasr{\"a}{\"a}si{\"o}, Skyt{\'e}n, J{\"a}rvenp{\"a}{\"a}, Marttinen,
  Gutmann, Vehtari, Corander, \& Kaski}]{elfi2018}
Lintusaari, J., Vuollekoski, H., Kangasr{\"a}{\"a}si{\"o}, A., {et~al.} 2018,
  Journal of Machine Learning Research, 19, 1.
\newblock \url{http://jmlr.org/papers/v19/17-374.html}

\bibitem[{MacDonald \& Petit(2019)}]{MacDonald2019}
MacDonald, J., \& Petit, V. 2019, Monthly Notices of the Royal Astronomical
  Society, 487, 3904, \dodoi{10.1093/mnras/stz1545}

\bibitem[{{Moe} \& {Di Stefano}(2017)}]{Moe:2017}
{Moe}, M., \& {Di Stefano}, R. 2017, \apjs, 230, 15,
  \dodoi{10.3847/1538-4365/aa6fb6}

\bibitem[{{Oda} {et~al.}(1994){Oda}, {Hino}, {Muto}, {Takahara}, \&
  {Sato}}]{Oda1994}
{Oda}, T., {Hino}, M., {Muto}, K., {Takahara}, M., \& {Sato}, K. 1994, Atomic
  Data and Nuclear Data Tables, 56, 231, \dodoi{10.1006/adnd.1994.1007}

\bibitem[{{Paxton} {et~al.}(2011){Paxton}, {Bildsten}, {Dotter}, {Herwig},
  {Lesaffre}, \& {Timmes}}]{Paxton2011}
{Paxton}, B., {Bildsten}, L., {Dotter}, A., {et~al.} 2011, \apjs, 192, 3,
  \dodoi{10.1088/0067-0049/192/1/3}

\bibitem[{{Paxton} {et~al.}(2013){Paxton}, {Cantiello}, {Arras}, {Bildsten},
  {Brown}, {Dotter}, {Mankovich}, {Montgomery}, {Stello}, {Timmes}, \&
  {Townsend}}]{Paxton2013}
{Paxton}, B., {Cantiello}, M., {Arras}, P., {et~al.} 2013, \apjs, 208, 4,
  \dodoi{10.1088/0067-0049/208/1/4}

\bibitem[{{Paxton} {et~al.}(2015){Paxton}, {Marchant}, {Schwab}, {Bauer},
  {Bildsten}, {Cantiello}, {Dessart}, {Farmer}, {Hu}, {Langer}, {Townsend},
  {Townsley}, \& {Timmes}}]{Paxton2015}
{Paxton}, B., {Marchant}, P., {Schwab}, J., {et~al.} 2015, \apjs, 220, 15,
  \dodoi{10.1088/0067-0049/220/1/15}

\bibitem[{{Paxton} {et~al.}(2018){Paxton}, {Schwab}, {Bauer}, {Bildsten},
  {Blinnikov}, {Duffell}, {Farmer}, {Goldberg}, {Marchant}, {Sorokina},
  {Thoul}, {Townsend}, \& {Timmes}}]{Paxton2018}
{Paxton}, B., {Schwab}, J., {Bauer}, E.~B., {et~al.} 2018, \apjs, 234, 34,
  \dodoi{10.3847/1538-4365/aaa5a8}

\bibitem[{{Paxton} {et~al.}(2019){Paxton}, {Smolec}, {Schwab}, {Gautschy},
  {Bildsten}, {Cantiello}, {Dotter}, {Farmer}, {Goldberg}, {Jermyn}, {Kanbur},
  {Marchant}, {Thoul}, {Townsend}, {Wolf}, {Zhang}, \& {Timmes}}]{Paxton2019}
{Paxton}, B., {Smolec}, R., {Schwab}, J., {et~al.} 2019, \apjs, 243, 10,
  \dodoi{10.3847/1538-4365/ab2241}

\bibitem[{{Potekhin} \& {Chabrier}(2010)}]{Potekhin2010}
{Potekhin}, A.~Y., \& {Chabrier}, G. 2010, Contributions to Plasma Physics, 50,
  82, \dodoi{10.1002/ctpp.201010017}

\bibitem[{{Rogers} \& {Nayfonov}(2002)}]{Rogers2002}
{Rogers}, F.~J., \& {Nayfonov}, A. 2002, \apj, 576, 1064,
  \dodoi{10.1086/341894}

\bibitem[{{Saumon} {et~al.}(1995){Saumon}, {Chabrier}, \& {van
  Horn}}]{Saumon1995}
{Saumon}, D., {Chabrier}, G., \& {van Horn}, H.~M. 1995, \apjs, 99, 713,
  \dodoi{10.1086/192204}

\bibitem[{{Schneider} {et~al.}(2019{\natexlab{a}}){Schneider}, {Ohlmann},
  {Podsiadlowski}, {R{\"o}pke}, {Balbus}, {Pakmor}, \&
  {Springel}}]{Schneider:19}
{Schneider}, F. R.~N., {Ohlmann}, S.~T., {Podsiadlowski}, P., {et~al.}
  2019{\natexlab{a}}, \nat, 574, 211, \dodoi{10.1038/s41586-019-1621-5}

\bibitem[{{Schneider} {et~al.}(2019{\natexlab{b}}){Schneider}, {Ohlmann},
  {Podsiadlowski}, {R{\"o}pke}, {Balbus}, {Pakmor}, \&
  {Springel}}]{Schneider:2019}
---. 2019{\natexlab{b}}, \nat, 574, 211, \dodoi{10.1038/s41586-019-1621-5}

\bibitem[{{Schneider} {et~al.}(2016){Schneider}, {Podsiadlowski}, {Langer},
  {Castro}, \& {Fossati}}]{Schneider:16}
{Schneider}, F.~R.~N., {Podsiadlowski}, P., {Langer}, N., {Castro}, N., \&
  {Fossati}, L. 2016, \mnras, 457, 2355, \dodoi{10.1093/mnras/stw148}

\bibitem[{{Sikora} {et~al.}(2019{\natexlab{a}}){Sikora}, {Wade}, {Power}, \&
  {Neiner}}]{Sikora2019}
{Sikora}, J., {Wade}, G.~A., {Power}, J., \& {Neiner}, C. 2019{\natexlab{a}},
  \mnras, 483, 2300, \dodoi{10.1093/mnras/sty3105}

\bibitem[{{Sikora} {et~al.}(2019{\natexlab{b}}){Sikora}, {Wade}, {Power}, \&
  {Neiner}}]{Sikora2019a}
---. 2019{\natexlab{b}}, \mnras, 483, 3127, \dodoi{10.1093/mnras/sty2895}

\bibitem[{{Spruit}(2002{\natexlab{a}})}]{Spruit2002}
{Spruit}, H.~C. 2002{\natexlab{a}}, \aap, 381, 923,
  \dodoi{10.1051/0004-6361:20011465}

\bibitem[{{Spruit}(2002{\natexlab{b}})}]{2002A&A...381..923S}
---. 2002{\natexlab{b}}, \aap, 381, 923, \dodoi{10.1051/0004-6361:20011465}

\bibitem[{{Stello} {et~al.}(2016){Stello}, {Cantiello}, {Fuller}, {Huber},
  {Garc{\'\i}a}, {Bedding}, {Bildsten}, \& {Silva Aguirre}}]{Stello:2016}
{Stello}, D., {Cantiello}, M., {Fuller}, J., {et~al.} 2016, \nat, 529, 364,
  \dodoi{10.1038/nature16171}

\bibitem[{Sunnåker {et~al.}(2013)Sunnåker, Busetto, Numminen, Corander, Foll,
  \& Dessimoz}]{sunnaker2013}
Sunnåker, M., Busetto, A.~G., Numminen, E., {et~al.} 2013, PLOS Computational
  Biology, 9, 1, \dodoi{10.1371/journal.pcbi.1002803}

\bibitem[{{Timmes} \& {Swesty}(2000)}]{Timmes2000}
{Timmes}, F.~X., \& {Swesty}, F.~D. 2000, \apjs, 126, 501,
  \dodoi{10.1086/313304}

\bibitem[{{Troland} \& {Crutcher}(2008)}]{2008ApJ...680..457T}
{Troland}, T.~H., \& {Crutcher}, R.~M. 2008, \apj, 680, 457,
  \dodoi{10.1086/587546}

\bibitem[{{Ud-Doula} {et~al.}(2009){Ud-Doula}, {Owocki}, \&
  {Townsend}}]{Ud-Doula2009}
{Ud-Doula}, A., {Owocki}, S.~P., \& {Townsend}, R. H.~D. 2009, \mnras, 392,
  1022, \dodoi{10.1111/j.1365-2966.2008.14134.x}

\bibitem[{{Vidotto} {et~al.}(2013){Vidotto}, {Jardine}, {Morin}, {Donati},
  {Lang}, \& {Russell}}]{Vidotto2013}
{Vidotto}, A.~A., {Jardine}, M., {Morin}, J., {et~al.} 2013, \aap, 557, A67,
  \dodoi{10.1051/0004-6361/201321504}

\bibitem[{{Villebrun} {et~al.}(2019){Villebrun}, {Alecian}, {Hussain},
  {Bouvier}, {Folsom}, {Lebreton}, {Amard}, {Charbonnel}, {Gallet},
  {Haemmerl{\'e}}, {B{\"o}hm}, {Johns-Krull}, {Kochukhov}, {Marsden}, {Morin},
  \& {Petit}}]{2019A&A...622A..72V}
{Villebrun}, F., {Alecian}, E., {Hussain}, G., {et~al.} 2019, \aap, 622, A72,
  \dodoi{10.1051/0004-6361/201833545}

\bibitem[{{Weber} \& {Davis}(1967)}]{Weber1967}
{Weber}, E.~J., \& {Davis}, Leverett, J. 1967, \apj, 148, 217,
  \dodoi{10.1086/149138}

\bibitem[{{Woltjer}(1964)}]{Woltjer1964}
{Woltjer}, L. 1964, \apj, 140, 1309, \dodoi{10.1086/148028}

\end{thebibliography}
